\documentclass{blois}
\usepackage{amsmath,amssymb}

\def\be{\begin{equation}}
\def\ee{\end{equation}}
\def\bea{\begin{eqnarray}}
\def\eea{\end{eqnarray}}

\newcommand{\Photo}

\begin{document}
\vspace*{4cm}
\title{Dark Matter Search 
with Cherenkov Telescope Array}

\author{Nagisa Hiroshima}

\address{Department of Physics, University of Toyama\\  3190 Gofuku, Toyama 930-8555, Japan,\\
RIKEN Interdisciplinary Theoretical and Mathematical Sciences (iTHEMS),\\
Wako, Saitama 351-0198, Japan}

\maketitle
\abstract{Many models of dark matter (DM) are now widely considered and probed intensively with accelerators, underground detectors, and astrophysical experiments. Among the various approaches, high-energy astrophysical observations are extremely useful to complement laboratory searches for some DM candidates. In the near future, the Cherenkov Telescope Array (CTA) should enable us to access much heavier weakly interacting massive particles, as well as a broad range of other DM candidates. In this talk, we describe DM searches with CTA.}


\section{Introduction}
\label{s:intro}
The nature of dark matter (DM), which accounts for about a quarter of the total energy density of the Universe~\cite{Planck:2018vyg} is still a mystery. Its existence is clear from cosmological and astrophysical observations on various scales, while the characterization of DM in the context of the particle physics models has not been achieved yet. Varieties of models are proposed: weakly interacting massive particles (WIMPs) and axion or axion-like particles (ALPs) are parts of the examples. Non-particle solutions such as primordial black holes are also widely discussed. 

Among the various candidates, WIMPs are one of the most promising ones. WIMP shares the same thermal bath with the Standard Model particle in the early Universe. The pair annihilation of WIMPs to the Standard Model particles is frequent enough to sustain the thermal equilibrium in the early Universe. It decouples from the thermal bath when the expansion rate of the Universe gets larger than the interaction rate. 
Particles of the weak-scale mass can satisfy the relic abundance if the thermally-averaged annihilation cross-section takes the canonical value $\langle \sigma v\rangle\sim{\cal O}(10^{-26}\ {\rm cm}^3/s)$~\cite{Steigman:2012nb,Saikawa:2020swg}, which also corresponds to the weak-scale value. The coincidence of the mass and the cross-section is dubbed as the WIMP miracle. 

The same process of WIMP annihilation to the Standard Model particles in the early Universe can be expected in the current Universe. High-energy $\gamma$-ray observations are powerful in probing such signatures. For example, non-detection of $\gamma$-rays from dwarf spheroidal galaxies (dSphs) by Fermi-LAT already excludes WIMPs of $m\lesssim{\cal O}(100{\rm GeV})$\cite{Hoof:2018hyn}. For WIMPs of $m\gtrsim{\cal O}(1{\rm TeV})$, observations by imaging atmospheric Cherenkov telescopes probe significant portions of the parameter space. In the near future, the Cherenkov Telescope Array (CTA) is expected to fill the search through the remaining parameter spaces and detect or exclude canonical WIMP models.

In this talk, we overview the forecasts for probing WIMP with CTA, introducing several key targets for observations. Prospects for other DM models are also discussed. 

\section{WIMP searches with CTA}
\label{s:target}
CTA is sensitive to very-high-energy $\gamma$-rays of $E_\gamma\sim{\cal O}(10{\rm GeV}-100{\rm TeV})$. High-energy photons entering the atmosphere generate cascades of particles, which are referred to as electromagnetic showers. The Cherenenkov image of the shower is measured by arrays of telescopes on the ground. The energy and direction of the initial high-energy photon are reconstructed from the image with the energy resolution $\Delta E/E\lesssim{\cal O}(0.1)$ and the angular resolution better than $\Delta\theta\lesssim{\cal O}(0.1^\circ)$ for $E_\gamma\gtrsim{\cal O}(100{\rm GeV})$. CTA will cover the whole sky by hosting arrays in both the Northern and Southern hemispheres~\footnote{https://www.cta-observatory.org/science/ctao-performance/}.

The $\gamma$-ray flux from WIMP annihilation $\phi_\gamma$, is expressed with annihilation spectrum $dN/dE$: 
\begin{equation}
    \phi_\gamma=\frac{1}{8\pi}\frac{\langle\sigma v\rangle}{m^2}\int dE \frac{dN}{dE}\int_{\Delta\Omega} d\Omega\int_{los} ds\rho_{\rm DM}^2
    \label{eq:flux}
\end{equation}
where $\int_{\Delta\Omega}d\Omega\int_{los}ds\rho_{\rm DM}^2$ is the so-called J-factor, which is the line-of-sight integral of the squared mass density of DM, $\rho_{\rm DM}$.  As it is seen from Eq.~\ref{eq:flux}, regions of high J-factors are promising for detecting WIMP annihilation signals.

CTA lists the following four targets for its key science project\footnote{CTA will operate as an open-proposal observatory. The key science projects consist of the prepared proposals by CTA which cover highly motivated science topics.~\cite{CTAConsortium:2017dvg}}:
\begin{itemize}
    \item the Galactic Center (GC),
    \item dwarf spheroidal galaxies (dSphs),
    \item Large Magellanic Cloud (LMC) and nearby galaxies,
    \item galaxy clusters.
\end{itemize}

\subsection{The Galactic Center (GC)}
\label{ss:GC}
The Galactic Center (GC) is the highest J-factor region and has been probed with various experiments~\cite{HESS:2018cbt,MAGIC:2022acl}. WIMP annihilating signatures may have been already visible with current facilities, however, it is still difficult to claim the detection~\cite{Fermi-LAT:2017opo,HESS:2006zwn}. The uncertainty in the modeling of the DM density distributions around the GC is one major issue in evaluating the annihilation flux. 
For obtaining the prospects with CTA, we have modeled the localized sources, interstellar $\gamma$-ray emissions due to the cosmic-ray interactions, and contributions from Fermi Bubbles, in addition to the residual cosmic-rays~\cite{CTA:2020qlo}. Comparison between Einasto~\cite{Einasto:1965} and generalized Navarro-Frenk-White profile~\cite{Navarro:1995iw,McMillan:2011wd} is also conducted. For example, the J-factor in the solid angle of 0.037~str around the GC is $7.1\times10^{22}$ GeV$^2$/cm$^5$ when the Einasto profile is assumed. Assuming this J-factor and 525 hours of observation with CTA, the sensitivity reaches the canonical cross-section for WIMPs of $m\sim0.3-10$~TeV ($0.2-20$~TeV) annihilating to $b\bar{b}$ ($W^-W^+$).

\subsection{Dwarf spheroidal galaxies (dSphs)}
Possibilities with dwarf spheroidal galaxies (dSphs) are also investigated in detail. They are expected to have high J-factors of $\lesssim{\cal O}(10^{19}{\rm GeV}^2/{\rm cm}^5)$ and do not show star formation activities. Hence the detection of $\gamma$-ray emission should be a promising WIMP annihilation signature. Up to now, no $\gamma$-ray emissions are confirmed for dSphs putting tight constraints on WIMP below $m\sim{\cal O}(100{\rm GeV})$. Continuous efforts to probe heavier WIMPs with this type of object by various experiments are further tightening the constraints~\cite{Fermi-LAT:2019lyf,Hess:2021cdp}. Our sensitivity with dSphs are limited by the understanding of the DM distribution. 500 hours of observations for one promising galaxy, for example Draco dSph, can probe down to $\langle\sigma v\rangle\sim10^{-24}-10^{-23}$~cm$^3$/s depending on the assumed profile~\cite{Hiroshima:2019wvj}. Further reductions of the systematic uncertainties are now on-going~\cite{CTAConsortium:2023nfo}.

\subsection{Large Magellanic Cloud (LMC) and nearby galaxies}
Large Magellanic Cloud (LMC) and nearby galaxies are advantageous in terms of J-factor due to their closeness. Prospects with nearby galaxies are derived taking M31 and M33 as examples~\cite{Michailidis:2023pkd}. A certain amount of observation time of CTA is planned to be assigned for those objects. Since they are massive and close to our Galaxy, spatial extension and substructures of DM halos need to be carefully modeled. Four and six sources are already found for regions of interest around M33 and M31, respectively. M31 also has extended emissions. Those emissions and residual cosmic-ray background are considered in the current modeling. For WIMP of $m\sim1$ TeV annihilating to $b\bar{b}$ ($\tau^-\tau^+$) pairs, cross-section down to $\langle \sigma v\rangle\sim 2\times 10^{-24} (10^{-24})$~cm$^3$/s and $\sim3\times10^{-23}(10^{-23})$~cm$^3$/s can be probed with M31 and M33, respectively. Depending on the assumption of the density profile and subhalo contributions, the constraints could get weaker.

\subsection{Galaxy clusters}
Galaxy clusters, such as the Perseus, reside in the largest-scale DM halos of $M\lesssim{\cal O}(10^{16})M_\odot$, i.e., significant amounts of DM are contained in such structures. They host lots of DM substructures and are promising targets, especially for decaying DM, to which sensitivity is controlled by the total mass and distance to the object rather than its concentration. Forecasts with CTA are obtained assuming 300 hours of observation of the Perseus cluster~\cite{Perez-Romero:2021gxh,CherenkovTelescopeArrayConsortium:2023niy}. This cluster is the brightest in X-ray observations and its astrophysical properties are well-investigated already. It hosts two active galactic nuclei as variable sources. Contributions from cosmic-ray-induced $\gamma$-ray emission are also modeled in the analysis. The annihilation cross-section can be probed down to $\langle\sigma v\rangle\sim10^{-23}$~GeV$^2$/s for $m~\sim 1$TeV in either annihilation channel of $b\bar{b}$, $W^-W^+$, and $\tau^-\tau^+$. The lifetime of the decaying DM can be constrained to $\tau\gtrsim10^{26}$s for the same DM mass.
\section{Beyond WIMP targets}
\label{s:ALP}
WIMP is not only the candidates we can probe with CTA. As a showcase for another well-motivated model, prospects for axion-like particles (ALPs) are also carefully evaluated~\cite{CTA:2020hii}. Astrophysical limits for such particles are obtained by searching the spectrum modulation induced by ALP($a$)-photon($\gamma$) conversion, the process of $\gamma+B\to a+B\to \gamma'+B\to\dots$ under background magnetic field $B$. Analysis with NGC1275, which resides at the center of the Perseus, gives the tightest constraints in the mass range of $m_a\sim10^{-10}-10^{-8}$~eV\cite{Fermi-LAT:2016nkz}. CTA probes a slightly larger mass region with the same objects. The quiescent(flare) spectrum of NGC1275 is modeled with a single power-law (with exponential cut-off) and wiggly features are searched in simulated 300(50) hours of observations. ALP-photon coupling can be probed down to $3\times10^{-12}$GeV$^{-1}$ assuming magnetic field strengh of $B=10\mu$G. The constraints depend on the assumption of the strength and the index of the turbulence of the background magnetic field.
\section{Summary}
\label{s:summary}
CTA, the next-generation very-high-energy $\gamma$-ray observatory now being constructed and starting its initial data taking, should powerfully probe varieties of DM candidates. 
WIMP is a prioritized candidate for CTA and the sensitivity down to the canonical cross-section can be achieved by observing the Galactic Center. Constraints expected with other objects are milder, however, they should surpass the ones obtained with current facilities. It is crucially important to update the accuracy of the modeling of our targets. Currently, the uncertainties are dominated by those in the DM density distributions and careful treatments of the astrophysical emissions are also required.  For ALP particles in the range of $m_a\sim{\cal O}(10^{-8})$ eV, coupling strength to the photon can be improved by about one order of magnitude. Our understanding of our Universe and physics beyond the Standard Model are to be deepened with CTA.


\section*{References}

\begin{thebibliography}{99}
\bibitem{Planck:2018vyg}
N.~Aghanim \textit{et al.} [Planck],
Astron. Astrophys. \textbf{641} (2020), A6
[erratum: Astron. Astrophys. \textbf{652} (2021), C4]
doi:10.1051/0004-6361/201833910
[arXiv:1807.06209 [astro-ph.CO]].

\bibitem{Steigman:2012nb}
G.~Steigman, B.~Dasgupta and J.~F.~Beacom,
Phys. Rev. D \textbf{86} (2012), 023506
doi:10.1103/PhysRevD.86.023506
[arXiv:1204.3622 [hep-ph]].

\bibitem{Saikawa:2020swg}
K.~Saikawa and S.~Shirai,
JCAP \textbf{08} (2020), 011
doi:10.1088/1475-7516/2020/08/011
[arXiv:2005.03544 [hep-ph]].

\bibitem{Hoof:2018hyn}
S.~Hoof, A.~Geringer-Sameth and R.~Trotta,
JCAP \textbf{02} (2020), 012
doi:10.1088/1475-7516/2020/02/012
[arXiv:1812.06986 [astro-ph.CO]].

\bibitem{CTAConsortium:2017dvg}
B.~S.~Acharya \textit{et al.} [CTA Consortium],
WSP, 2018,
ISBN 978-981-327-008-4
doi:10.1142/10986
[arXiv:1709.07997 [astro-ph.IM]].

\bibitem{HESS:2018cbt}
H.~Abdallah \textit{et al.} [HESS],
Phys. Rev. Lett. \textbf{120} (2018) no.20, 201101
doi:10.1103/PhysRevLett.120.201101
[arXiv:1805.05741 [astro-ph.HE]].

\bibitem{HESS:2006zwn}
F.~Aharonian \textit{et al.} [H.E.S.S.],
Phys. Rev. Lett. \textbf{97} (2006), 221102
[erratum: Phys. Rev. Lett. \textbf{97} (2006), 249901]
doi:10.1103/PhysRevLett.97.221102
[arXiv:astro-ph/0610509 [astro-ph]].


\bibitem{MAGIC:2022acl}
H.~Abe \textit{et al.} [MAGIC],
Phys. Rev. Lett. \textbf{130} (2023) no.6, 061002
doi:10.1103/PhysRevLett.130.061002
[arXiv:2212.10527 [astro-ph.HE]].

\bibitem{Fermi-LAT:2017opo}
M.~Ackermann \textit{et al.} [Fermi-LAT],
Astrophys. J. \textbf{840} (2017) no.1, 43
doi:10.3847/1538-4357/aa6cab
[arXiv:1704.03910 [astro-ph.HE]].



\bibitem{CTA:2020qlo}
A.~Acharyya \textit{et al.} [CTA],
JCAP \textbf{01} (2021), 057
doi:10.1088/1475-7516/2021/01/057
[arXiv:2007.16129 [astro-ph.HE]].

\bibitem{Einasto:1965}
J.~Einasto,
Trudy Astrofizicheskogo Instituta Alma-Ata 5, 87-100

\bibitem{Navarro:1995iw}
J.~F.~Navarro, C.~S.~Frenk and S.~D.~M.~White,
Astrophys. J. \textbf{462} (1996), 563-575
doi:10.1086/177173
[arXiv:astro-ph/9508025 [astro-ph]].
\bibitem{McMillan:2011wd}
P.~J.~McMillan,
Mon. Not. Roy. Astron. Soc. \textbf{414} (2011), 2446-2457
doi:10.1111/j.1365-2966.2011.18564.x
[arXiv:1102.4340 [astro-ph.GA]].

\bibitem{Fermi-LAT:2019lyf}
L.~Oakes \textit{et al.} [Fermi-LAT, HAWC, H.E.S.S., MAGIC and VERITAS],
PoS \textbf{ICRC2019} (2021), 012
doi:10.22323/1.358.0012
[arXiv:1909.06310 [astro-ph.HE]].

\bibitem{Hess:2021cdp}
H.~Abdalla \textit{et al.} [Hess, HAWC, VERITAS, MAGIC, H.E.S.S. and Fermi-LAT],
PoS \textbf{ICRC2021} (2021), 528
doi:10.22323/1.395.0528
[arXiv:2108.13646 [hep-ex]].


\bibitem{Hiroshima:2019wvj}
N.~Hiroshima, M.~Hayashida and K.~Kohri,
Phys. Rev. D \textbf{99} (2019) no.12, 123017
doi:10.1103/PhysRevD.99.123017
[arXiv:1905.12940 [astro-ph.HE]].

\bibitem{CTAConsortium:2023nfo}
F.~G.~Saturni \textit{et al.} [CTA Consortium],
PoS \textbf{ICRC2023} (2023), 1366
doi:10.22323/1.444.1366


\bibitem{Perez-Romero:2021gxh}
J.~P\'erez-Romero [CTA Consortium],
PoS \textbf{ICRC2021} (2021), 546
doi:10.22323/1.395.0546
[arXiv:2108.05141 [astro-ph.HE]].

\bibitem{CherenkovTelescopeArrayConsortium:2023niy}
K.~Abe \textit{et al.} [Cherenkov Telescope Array Consortium],
[arXiv:2309.03712 [astro-ph.HE]].

\bibitem{Michailidis:2023pkd}
M.~Michailidis, L.~Marafatto, D.~Malyshev, F.~Iocco, G.~Zaharijas, O.~Sergijenko, M.~I.~Bernardos, C.~Eckner, A.~Boyarsky and A.~Sokolenko, \textit{et al.}
[arXiv:2304.08202 [astro-ph.HE]].



\bibitem{CTA:2020hii}
H.~Abdalla \textit{et al.} [CTA],
JCAP \textbf{02} (2021), 048
doi:10.1088/1475-7516/2021/02/048
[arXiv:2010.01349 [astro-ph.HE]].

\bibitem{Fermi-LAT:2016nkz}
M.~Ajello \textit{et al.} [Fermi-LAT],
Phys. Rev. Lett. \textbf{116} (2016) no.16, 161101
doi:10.1103/PhysRevLett.116.161101
[arXiv:1603.06978 [astro-ph.HE]].





\end{thebibliography}
\end{document}